\documentstyle[sprocl]{article}

\input{psfig.sty}

\def\be{\begin{equation}}
\def\ee{\end{equation}}
\def\bea{\begin{eqnarray}}
\def\eea{\end{eqnarray}}
\begin{document}

\title{CHARACTERIZATION OF UNIVERSAL BEHAVIOR \\IN QCD DIRAC SPECTRA}

\author{J.J.M. VERBAARSCHOT}

\address{Department of Physics and Astronomy,\\
         University at Stony Brook,\\
         Stony Brook, NY\,11794, USA}   


\maketitle\abstracts{In this lecture we discuss correlations of the
QCD Dirac eigenvalues. We find that below a scale of $E_c\sim \Lambda/L^2$ they
are given by chiral Random Matrix Theory. This follows from 
analytical arguments based on 
partially quenched Chiral Perturbation Theory and is 
substantiated by lattice QCD and instanton liquid simulations. }

\section{INTRODUCTION}

The low-energy limit of QCD 
is characterized by two parameters, the chiral condensate, $\Sigma$, and the
pion decay constant, $F$. The chiral condensate is directly 
related to the density
of the smallest eigenvalues of the Dirac operator by means of the Banks-Casher 
formula \cite{BC}. At finite volume, the pion decay constant determines
the quark mass scale below which the QCD partition function is given 
by the zero momentum component of the chiral Lagrangian \cite{GL,LS}.
Below this scale the Dirac eigenvalues are constraint by Leutwyler-Smilga
sum rules~\cite{LS}. On the other hand, the accumulation of small
eigenvalues for broken chiral symmetry 
requires strong interactions \cite{cam97}.  From the study of
complex systems (see Guhr et al. \cite{HDgang}
for a comprehensive review) we expect that microscopic 
correlations of eigenvalues of such disordered system
are universal and  can be described by the simplest model in its
universality class. Such
a model is chiral Random Matrix Theory (chRMT). The interpretation is
that the classical motion of quarks in Euclidean QCD with an additional
artificial time dimension is chaotic.

These ideas can be formulated more precisely in terms of partially quenched
Chiral Perturbation Theory~\cite{pqChPT} (pqChPT), 
which, in addition to the usual quarks, contains valence quarks and
their  superpartners \cite{Morel}. If both the Vafa-Witten
theorem \cite{Vafa-Witten} 
and the Goldstone theorem apply to this theory, only
  one mass term consistent
with the chiral symmetries of the partition function can be written down. 
On the other hand, it can be shown \cite{Toublan}
 that the zero-momentum sector of pqChPT reproduces identically the chRMT 
distribution of the smallest eigenvalues up to a scale 
which is the equivalent of the Thouless energy in the theory of mesoscopic 
systems \cite{Altshuler,HDgang}. In
units of the level spacing, this scale is given by $n_c = F^2 L^2/\pi$
for a box of length $L$.
This argument proves the conjecture \cite{SVR}  
that the correlations of the smallest QCD Dirac eigenvalues are given
by chRMT. 
The stability of the RMT correlations under deformations of the 
ensemble as shown by universality
studies \cite{Brezin,Damgaard,GWu,Sener1,Dampart}
\cite{andystudent,Tilodam,Senerprl,Brezinmulti}
strengthens our confidence in this result.
These arguments have been substantiated both by
instanton liquid \cite{Vinst,Osborn} and lattice QCD
simulations~\cite{tiloprl,Ma,many,Guhrth,Tilomass,HV,HKV,Markum,Luo}.

Because of the Banks-Casher relation, $\Sigma=\pi \rho(0)/V$,  
the eigenvalues near zero are spaced as 
$1/\rho(0) = \pi/\Sigma V$. In order to study the approach to the thermodynamic
limit it is natural to introduce the microscopic limit in which $u=\lambda V 
\Sigma$ is kept fixed for $V\rightarrow \infty$ and the microscopic spectral
density \cite{SVR}
\be
\rho_S(u) = \lim_{V\rightarrow \infty} \frac 1{V\Sigma} \langle
\rho(\frac u{V\Sigma})\rangle.
\label{rhosu}
\ee
Our claim is that for $u \ll n_c$ the microscopic spectral 
density is given by chRMT. Indeed, the chRMT result for (\ref{rhosu}) 
can be derived from pqChPT \cite{Toublan}.

The chRMT partition function is introduced in section 2. Its domain of 
validity is discussed in section 3 and  universality arguments are presented
in section 4. Lattice results are shown in section 5. Before concluding we
finish with applications of chRMT to QCD at 
nonzero chemical potential (section 6).

\section{CHIRAL RANDOM MATRIX THEORY}

In this section we will introduce an instanton 
liquid \cite{shurrev,diakonov} inspired 
chiral RMT  with the global symmetries of the QCD partition function but
otherwise  Gaussian random matrix elements. For $N_f$ flavors in the sector
with topological charge $\nu$ such chRMT  is defined by \cite{SVR,V}
\be
Z_{N_f,\nu}^\beta(m_1,\cdots, m_{N_f}) = 
\int DW \prod_{f= 1}^{N_f} \det({\rm \cal D} +m_f)
e^{-\frac{N\Sigma^2 \beta}4 {\rm Tr}W^\dagger W},
\label{zrandom}
\ee
where
\be
{\cal D} = \left (\begin{array}{cc} 0 & iW\\
iW^\dagger & 0 \end{array} \right ),
\label{diracop}
\ee
and $W$ is a $n\times m$ matrix with $\nu = |n-m|$ and
$N= n+m$. The matrix elements of $W$ are either real ($\beta = 1$, chiral
Gaussian Orthogonal Ensemble (chGOE)), complex
($\beta = 2$, chiral Gaussian Unitary Ensemble (chGUE)),
or quaternion real ($\beta = 4$, chiral Gaussian Symplectic Ensemble (chGSE)).
As is the case in QCD, we assume that $\nu$ does not exceed $\sqrt N$, so that,
to a good approximation, $n = N/2$. The value of the Dyson index $\beta$ is
determined by the anti-unitary symmetries of the Dirac operator. If there are
no anti-unitary symmetries the value of $\beta = 2$. 
If the square of the anti-unitary
symmetry operator is ${\bf 1}$ we have $\beta =1$ and if its square is
${\bf -1}$ we have $\beta = 4$. Together with the Dyson ensembles, the
GOE, the GUE and the GSE, the chiral ensembles can be classified in terms
symmetric spaces \cite{class}.

In this model chiral symmetry is broken spontaneously with chiral condensate
given by  $\Sigma = \lim_{N\rightarrow \infty} {\pi \rho(0)}/N$, where
$N$ is interpreted as the (dimensionless) volume of space
time. For complex matrix elements ($\beta =2$), which is appropriate for QCD 
with three or more colors and fundamental fermions, the
symmetry breaking pattern is {\cite{SmV}} $SU(N_f) \times SU(N_f)/SU(N_f)$.
The average spectral density that can be derived from (\ref{zrandom}) 
has the familiar semi-circular shape.
The microscopic spectral density for the chGUE
is given by {\cite{VZ,V}}
\be
\rho_S(z) = \frac z2 \left ( J^2_{a}(z) -
J_{a+1}(z)J_{a-1}(z)\right),
\label{micro2}
\ee
where $a = N_f + |\nu|$. 
The spectral correlations in the bulk of the spectrum are given 
by the invariant random matrix ensembles \cite{Kahn,nagao}. 

\section{DOMAIN OF VALIDITY OF chRMT}

The domain of validity is best discussed within the context of 
partially quenched Chiral Perturbation Theory (pqChPT). 
This is an effective field
theory for the low-energy limit of a QCD like theory that 
in addition to $N_f$ sea quarks contains 
valence quarks and their superpartners. It allows us to calculate
the valence quark mass dependence of the chiral condensate defined as
\be
\Sigma_v(m_v) = \frac 1V\int d\lambda \langle\rho(\lambda)\rangle
\frac{2m_v}{\lambda^2 +m^2_v}.
\label{sigmam}
\ee
The spectral density follows from the discontinuity of
$\Sigma_v(m_v)$,
\be
\frac{2\pi}V \langle \rho(\lambda)\rangle
=
%\left .{\rm Disc}\right |_{m_v = i\lambda}\Sigma(m_v) \equiv 
\lim_{\epsilon \rightarrow 0}
\Sigma(i\lambda+\epsilon) - \Sigma(i\lambda-\epsilon) = \frac{2\pi}V \sum_k
\langle \delta(\lambda +\lambda_k)\rangle,
\label{spectdisc}
\ee
where the average $\langle \cdots \rangle $ is with respect to the
distribution of the eigenvalues. 
Similarly, the two-point spectral correlation function 
follows from the double discontinuity of the scalar susceptibility. 

%\be
%\langle \rho(\lambda) \rho(\lambda') \rangle
%= \frac 1{4\pi^2}\left . {\rm Disc}
%\right |_{m_v = i\lambda, m_{v'}=i\lambda'}
%  \sum_{k,l}\left\langle
% \frac 1{i\lambda_k +m_v}\frac 1{i\lambda_l +m_{v'}}\right\rangle.
%\ee
%The valence quark mass dependence on $m_v$ and $m_{v'}$ can be calculated
%from the pqChPT partition function.
In both cases we can identify an important scale where the inverse mass
of the Goldstone modes corresponding to the valence quark is equal to
the size of the box. Using the relation $M= (m + m')\Sigma/F^2$, where $F$ is
the pion decay constant, we find from $ML =1$ that in terms of $m$ 
this scale is given by \cite{Osborn,Toublan,Janik}
\be
E_c = \frac {F^2}{\Sigma L^2}.
\ee
For $m_v \ll E_c$ we have shown that the valence quark mass dependence 
is given by the chRMT. The asymptotic result for the chRMT 
two-point correlation function follows from the double Goldstone pole 
in the scalar susceptibility.
The conclusion is that if
pqChPT describes correctly the low energy limit of QCD we have shown that
the correlations of the Dirac eigenvalues close to zero are given by 
chRMT. 

Such picture is well-known from mesoscopic physics \cite{HDgang}. 
In this context 
$E_c$ is defined as the inverse tunneling time of an electron through
the sample and is given by $E_c = {\hbar D}/{L^2}$,
where $D$ is the diffusion constant. Another scale that enters in 
these systems is the elastic scattering time $\tau_e$. Based on these scales
one can distinguish three different regimes \cite{Altshuler}
for the energy difference, 
$\delta E$, that enters in the two-point correlation
function: i) the ergodic regime
for $\delta E \ll E_c$, ii) the diffusive domain for $E_c \ll \delta E \ll 
\hbar/\tau_e$  
and iii) the ballistic regime for $\delta_E \gg \hbar/\tau_e$ (not discussed 
below).
For time scales corresponding to the ergodic regime an initially localized
wave packet covers all of phase space. In this domain the eigenvalue
correlations are given by RMT. In the diffusive domain an initially localized
wave packet 
explores only  part of the phase space resulting in weaker correlations between
the eigenvalues.
For earlier applications of localization theory to the chiral phase transition
we refer to \cite{shuryak,Smilref}.

Based on these ideas we can interpret the Euclidean Dirac spectrum as the 
energy levels of a Hamiltonian conjugate to an additional 
artificial time dimension.
According to the Bohigas conjecture \cite{Bohigas} 
the eigenvalue correlations are given by
RMT if and only if the corresponding classical motion is chaotic. We thus
conclude  that the classical time evolution of quarks in the Yang-Mills
gauge fields is chaotic.

\begin{center}
\begin{figure}[!ht]
\hspace{2.5cm}
\hbox{\psfig{file=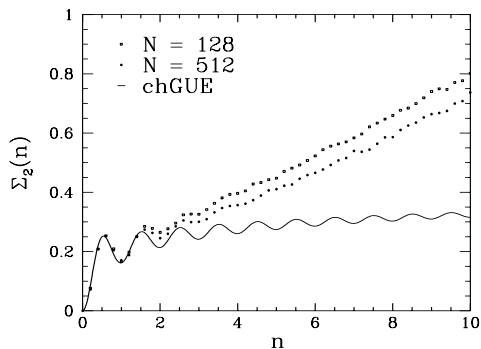,width=65mm,height=47mm}}
%\hbox{\psfig{file=sys$user:[osborn.plots]nv0a.ps,width=65mm,height=47mm}}
\caption{
The number variance $\Sigma_2(n)$ versus $n$ measured from an interval
starting at $\lambda = 0$. The total number of instantons is denoted by $N$.}
\label{fig1}
%\vspace*{-0.5cm}
\end{figure}
\end{center}
These ideas can be tested by means of lattice QCD \cite{many,Guhrth} 
 and instanton liquid~\cite{Osborn}
simulations. In Fig. 1, we show \cite{Osborn} $\Sigma_2(n)$, 
defined as the variance of
the number of levels in an interval containing $n$ level on average, versus 
$n$ for eigenvalues obtained
from the Dirac operator in the background of instanton liquid gauge
field configurations.
The chRMT result,  given by the solid curve, is reproduced 
up to about two level spacings.
In units of the  average level spacing, $\Delta=1/\rho(0)=\pi/\Sigma V$, 
the energy $E_c$ is given by
$n_c \equiv {E_c}/{\Delta} = {F^2 L^2}/\pi$.  
For an instanton liquid with instanton density $N/V =1$ we find that
$n_c \approx 0.07 \sqrt N$. We conclude that chRMT appears to 
describe the eigenvalue correlations up to the 
predicted scale.

The additional two flavors in  
Damgaard's relation \cite{Dampart} between the microscopic spectral density and 
the finite volume partition function arise naturally in pqChPT \cite{Toublan},
namely one valence flavor and its supersymmetric partner.

\section{UNIVERSALITY IN CHIRAL RANDOM MATRIX THEORY}

The aim of universality studies is to identify observables that are
stable against deformations of the 
random matrix ensemble. Not all observables have the same degree
of universality. For example, a semicircular average spectral density
is found for random matrix ensembles with independently distributed 
matrix elements with a finite variance. However, this
spectral shape does not occur in nature, and it is thus not
surprising that it is only found in a rather narrow class of random matrix
ensembles. What is surprising is that the $microscopic$ spectral density
and the $microscopic$ spectral correlators are stable with respect to
a much larger class of deformations.
Two different types of deformations have been considered, those that
maintain the unitary invariance of the partition functions and those
that break the unitary invariance.

In the  first class, the Gaussian probability
distribution is replaced by $
P(W) \sim \exp(-N \sum_{k=1}^\infty a_k {\rm Tr} 
(W^\dagger W)^k).$
For a potential with only $a_1$ and $a_2$ different from zero it was 
shown \cite{Brezin} that
the microscopic spectral density is independent of $a_2$. A general proof 
valid for arbitrary potential and all correlation functions
was given by 
Akemann et al. \cite{Damgaard}.
The essence of the proof
is a remarkable generalization of the identity for
the Laguerre polynomials, $\lim_{n \rightarrow \infty}  L_n(x/ n) =
 J_0(2 \sqrt x) \ , $
to orthogonal polynomials determined by an arbitrary potential. 
It was proved from the continuum limit of the recursion relation 
for orthogonal polynomials.

In  the second class, an arbritrary fixed matrix is added
to $W$ in the Dirac operator (\ref{diracop}).
It has been shown that the microscopic spectral density and 
the microscopic spectral correlations remain unaffected \cite{GWu,Sener1} 
for parameter values that completely modify the average spectral density.

Microscopic universality for deformations that affect the
macroscopic spectral density implies the existence of a scale 
beyond which universality breaks down. 
It can be interpreted naturally in terms of the
spreading width \cite{spreading}.

Based on the general form of of the pqChPT partition function one 
could argue that universality follows from its relation with chRMT. 
However, 
universality can be formulated as the stability of
the effective partition function with respect to variations of the 
distribution of matrix elements. The 
stability of the saddle-point manifold was demonstrated
for the Unitary ensembles \cite{Hack}.

\section{LATTICE QCD RESULTS}

In this section we consider correlations of lattice QCD Dirac eigenvalues.
For a Dirac operator with a $U_A(1)$ symmetry the eigenvalues 
occur in pairs $\pm \lambda_k$. Therefore  we have to distinguish
two different regions: the region near zero virtuality and the bulk of the
spectrum. The $U_A(1)$ symmetry is absent for the 
Hermitean Wilson Dirac operator which includes an additional $\gamma_5$ matrix.

The relevant class of random matrix ensembles is determined by 
the anti-unitary symmetries of the Dirac operator. For  
the $SU(2)$ color group,
the anti-unitary symmetries of the Kogut-Susskind (KS)
and the Wilson Dirac operator are given by {\cite{Teper,HV}},
\be
[D^{KS}, \tau_2 K] = 0, \quad {\rm and} \quad 
[\gamma_5 D^W, \gamma_5 CK\tau_2] = 0.
\ee
Because $(\tau_2 K)^2 = -1$ and $(\gamma_5 CK\tau_2)^2 =1$,
we have $\beta =1$ for Wilson fermions and $\beta = 4$ for KS fermions.
This results in the chGUE for KS fermions and, in absence of the
$U_A(1)$ symmetry, in the GOE for Wilson fermions \cite{HV}.
For three or more colors there are no anti-unitary symmetries and 
the relevant ensembles are the chGUE and GUE for KS and Wilson fermions,
respectively.

The eigenvalue correlations, which are separated from the 
average spectral density
by calculating them for the unfolded eigenvalues, 
are studied by means of moments of the number of
levels, $n_i$, in an interval  containing $n$ levels on average.
Below we discuss the variance, $\Sigma_2(n)$ and the first two cumulants
$\gamma_1(n)$ and $\gamma_2(n)$. 
The moments can be obtained either by ensemble averaging or by
spectral averaging. Under the assumption of spectral ergodicity, both
procedures should give the same results. However, recent results
\cite{Guhrth}
indicate that the domain of RMT correlations
is the much longer for spectral averaging than for ensemble 
averaging. This interesting observation certainly deserves more attention.

\begin{figure}[!ht]
%\vspace*{-1cm}
%\hspace*{1cm}
\vbox{
\hbox{
\psfig{file=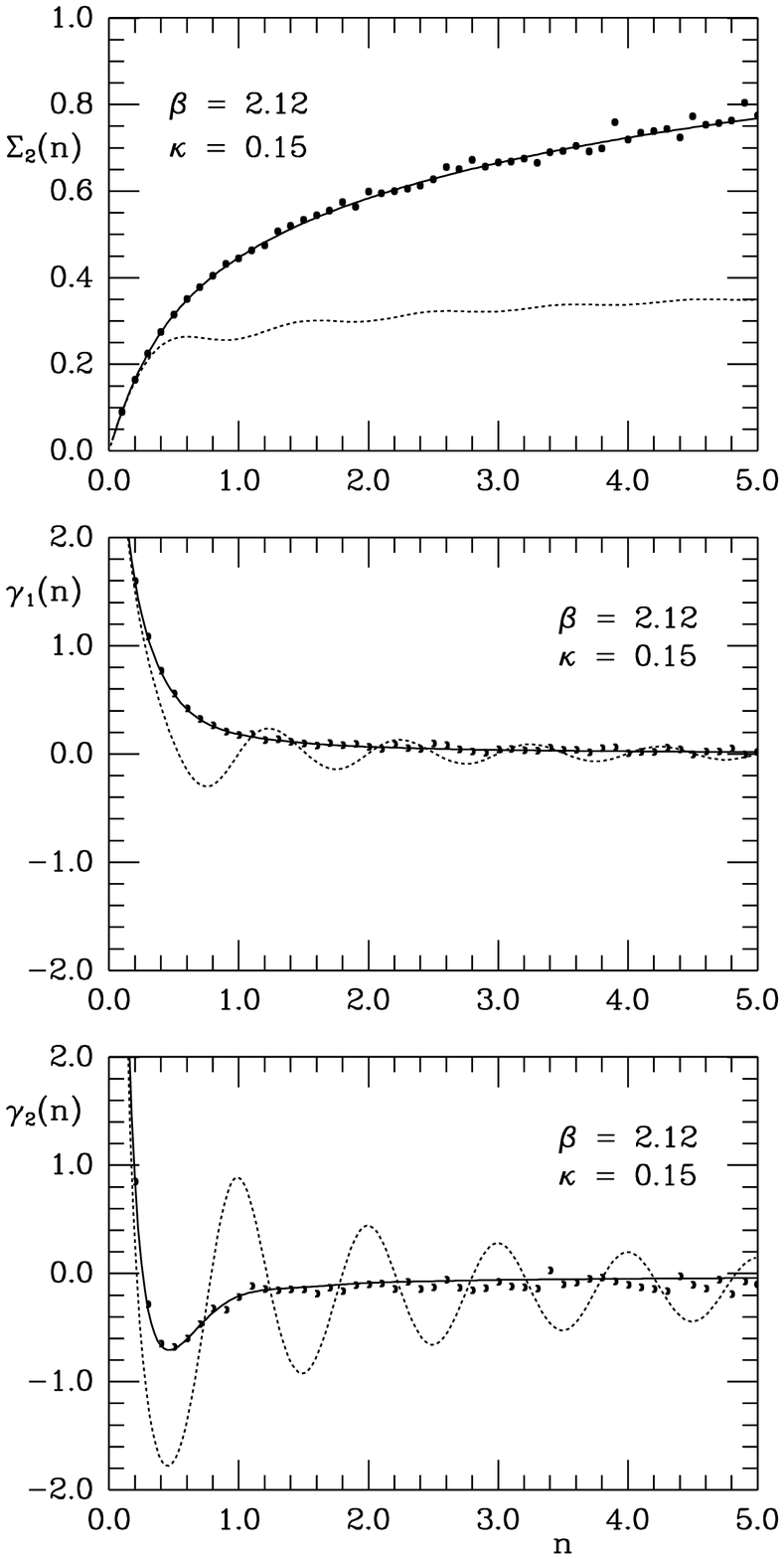,width=50mm}
\hspace*{1cm}
\psfig{file=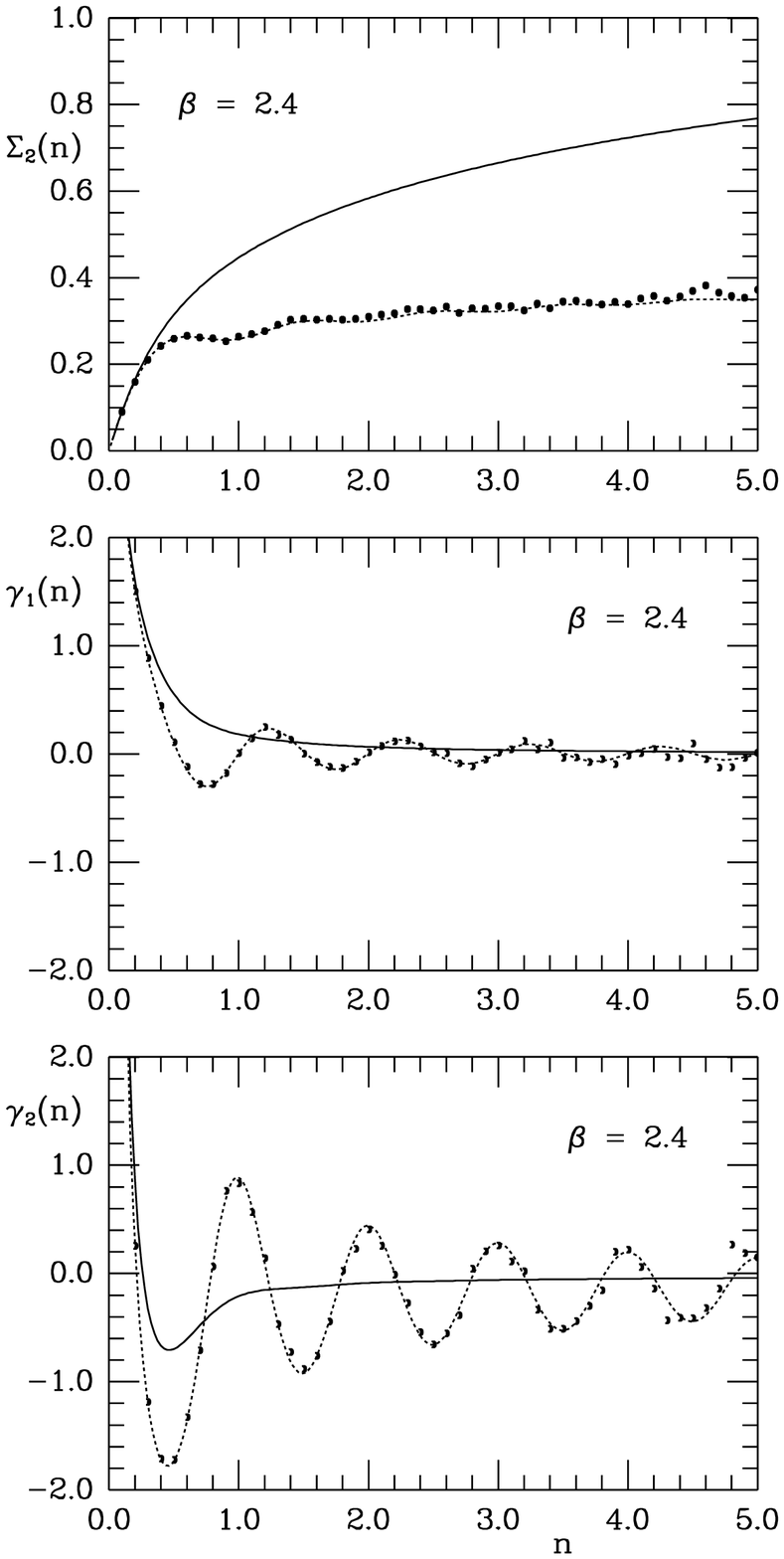,width=50mm}}}
\hspace*{1cm}
\vspace*{-0.25cm}
\caption{The number variance, $\Sigma_2(n)$, and the first two cumulants,
$\gamma_1(n)$ and $\gamma_2(n)$, as a function of $n$.}
% for  eigenvalues
%of the Wilson Dirac operator (left) and the Kogut-Susskind Dirac operator 
%(right). The full and dotted
%curves represent the analytical result for the GOE and the GSE, respectively.}
%\vspace*{-0.5cm}
\label{fig3}
\end{figure}
 
Results obtained by spectral averaging 
for $\Sigma_2(n)$, $\gamma_1(n)$ and $\gamma_2(n)$
for both KS and Wilson fermions with $N_c = 2$
(see Fig. \ref{fig3})
show an impressive agreement with the 
RMT predictions. This agreement extends to a wide range of $\beta$ values
ranging from strong coupling to weak coupling. 
The simulations for KS fermions were performed {\cite{Kalkreuter}}
for 4 dynamical flavors
with $ma = 0.05$ on a $12^4$ lattice. The simulations for Wilson fermions were
done for two dynamical flavors on a $8^3\times 12$ lattice. 
Recently, impressive results were obtained  for the nearest neighbor
spacing distribution of $N_c =3$ staggered Dirac spectra \cite{Markum}.

\begin{figure}[!ht]
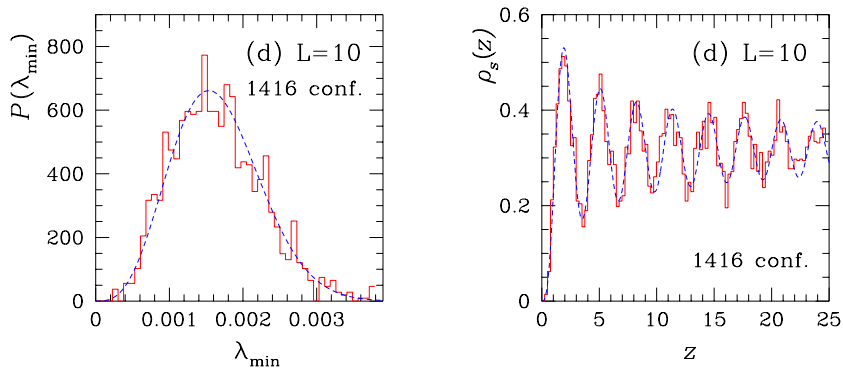

\vspace*{-1cm}
\hbox{
\psfig{file=tilo10small.ps,width=52mm,angle=0}
\hspace*{0.5cm}
\psfig{file=tilo10micro.ps,width=52mm,angle=0}}
%\psfig{file=[verbaarschot.zak]tilo10small.ps,width=52mm,angle=0}
%\hspace*{0.5cm}
%\psfig{file=[verbaarschot.zak]tilo10micro.ps,width=52mm,angle=0}}
\vspace*{-1.50cm}
\caption{The distribution of the smallest eigenvalue (left) and the
microscopic spectral density (right) 
for two colors and $\beta = 2.0$.} 
%Lattice results are represented by
%the histogram, and the analytical results for the chGSE are given by the
%dashed curves.}
\label{fig4}
%\vspace*{-0.5cm}
\end{figure}

Spectral ergodicity cannot
be exploited in the study of the microscopic spectral density.
In order to gather sufficient statistics,
a large number of independent spectra is required.
One way to proceed is to use instanton-liquid configurations
which can be generated cheaply. In this case we find
{\cite{Vinst}} that the microscopic spectral density is given
by the chGOE and the chGUE for $N_c=2$ and $N_c = 3$, respectively.
Recently, this analysis was performed for  lattice QCD Dirac eigenvalues
\cite{tiloprl}. Results for 1416 
 quenched $SU(2)$ Kogut-Susskind
Dirac spectra  on a 10$^4$ lattice are shown in Fig. \ref{fig4}. 
We show both the distribution of 
the smallest eigenvalue (left) and the microscopic spectral density (right).
The results for the chGSE are represented by the dashed curves.
\noindent
We  emphasize that the theoretical curves have been
obtained without any fitting of parameters. The input parameter, the
chiral condensate, is derived from the same lattice calculations. 
The above simulations were performed at a relatively strong coupling of
$\beta = 2$, but agreement with the chGSE predictions
was also found for  $\beta = 2.2$
and for $\beta =2.5$ on a $16^4$ lattice {\cite{tiloprl}}.

In the case of two fundamental colors the continuum theory 
and Wilson fermions are in the same universality class.
It is an interesting question of how spectral correlations of KS fermions
evolve in the approach to the continuum limit. Certainly, the 
Kramers degeneracy of the eigenvalues remains. However, since Kogut-Susskind
fermions represent 4 degenerate flavors in the continuum limit, 
the Dirac eigenvalues should obtain an additional two-fold degeneracy.
We are looking forward to more work in this direction.

\section{APPLICATIONS OF chRMT AT $\mu \ne  0$}

In the continuum formulation of QCD the chemical potential enters in the
QCD partition function by the addition of the term $\mu \gamma_0$ to the
anti-Hermitean Dirac operator, i.e. ${\cal D} \rightarrow {\cal D} + \mu 
\gamma_0$. In a suitable chiral
basis in which the matrix elements of $\langle 
\phi_R^k|\gamma_0|\phi_L^k\rangle= \delta_{kl}$, the modification
in the random matrix partition function  (\ref{zrandom})  
corresponds to replacing the Dirac matrix ${\cal D}$ by
\be
{\cal D} =
\left ( \begin{array}{cc} 0 & iW + \mu\\
iW^\dagger +\mu & 0 \end{array} \right) \  .
\label{Diracmatter}
\ee
The term $\mu\gamma_0$ has the same anti-unitary symmetries as
$\gamma_0\partial_0$ and therefore does not alter the corresponding 
classification of the Dirac operators.

%resulting in the partition function
%\be
%Z(m,\mu) = \int D W  
%{\det}^{N_f} {\cal D} e^{-n {\Sigma^2} {\rm Tr} WW^\dagger}
%\label{zmu}
%\ee

As is the case in QCD \cite{everybody}, 
a nonzero chemical potential violates the 
anti-Hermiticity of the Dirac operator and its eigenvalues are scattered in the
complex plane. 
One important result obtained from the chRMT at $\mu \ne 0$ is that
 the quenched approximation is the limit $N_f\rightarrow 0$ of the
partition function  with the determinant replaced by its
absolute value \cite{Stephanov}. 
By replacing $\mu \rightarrow \mu +iT$ one obtains 
a Landau-Ginzburg functional for the order parameter \cite{misha}.
A similar Landau-Ginzburg functional has been derived from the Nambu model 
\cite{krishna}.  For $\mu =0$ this
model undergoes a second order phase transition \cite{JV,Tilo,Stephanov1}. 
At $T= 0$ we find a first order phase transition \cite{Stephanov}. 
A tricritical point is found for at $m =0$, $\mu \ne 0$ and $T\ne 0$. 
For more discussion of this model we refer to the literature 
\cite{Feinberg-Zee}.

\begin{figure}[!ht]
%\hspace*{1cm}
\hbox{
\psfig{file=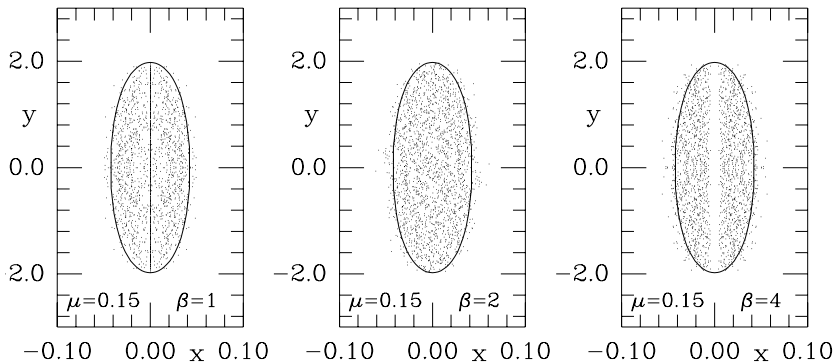,width=110mm}}
\caption{
Scatter plot of the real ($x$), and the imaginary
parts ($y$) of the eigenvalues of the random matrix Dirac operator
at $\mu \ne 0$.}
%The values of $\beta$ and $\mu$ are given in the labels of the figure.
%The full curve shows the analytical result for the boundary.}
\label{Fig6}
%\vspace*{-0.5cm}
\end{figure}

We close this lecture with identifying possible universal behavior 
at $\mu \ne 0$.
Quenched numerical simulations have been performed for all 
three classes of $\beta$.
A cut along the imaginary axis below
cloud of eigenvalues was found in instanton liquid
simula\-tions {\cite{Thomas}}  for $N_c =2$ at $\mu \ne 0$ ($\beta =1$). 
In lattice QCD simulations
with staggered fermions {\cite{baillie}} for $N_c = 2$ ($\beta= 4$) 
a depletion of eigenvalues along the imaginary axis was observed whereas
for $N_c=3$ ($\beta =2$), 
the eigenvalue distribution did not show any pronounced features.

In the quenched approximation, the spectral properties
of the random matrix ensemble (\ref{Diracmatter})
can be studied numerically by simply diagonalizing a set of
matrices with probability distribution (\ref{zrandom}).
In Fig. \ref{Fig6}  we show results {\cite{Osbornmu}} 
for the eigenvalues of a few
$100\times  100$ matrices for $\mu = 0.15$ (dots) of the quenched
random matrix ensemble (\ref{Diracmatter}).
The solid curve represents
the analytical result for the boundary of the domain of 
eigenvalues \cite{Stephanov,Osbornmu}.
We observe the accumulation and depletion that
was found in the previously mentioned simulations. 
This depletion can be understood as follows. For $\mu = 0$ all eigenvalues
are doubly degenerate. This degeneracy is broken at $\mu\ne 0$ which produces
the observed repulsion of the eigenvalues.

The number of purely imaginary eigenvalues for $\beta = 1$
appears to scale as $\sqrt N$  which has been understood analytically
for ensembles without a chiral structure {\cite{fyodorov}}. 
Obviously, more work has to be done in order to
arrive at a complete characterization of
universal features {\cite{fyodorovpoly}} of nonhermitean matrices.

\section{CONCLUSIONS}
We have presented both analytical and numerical arguments showing 
that the correlation
of the QCD Dirac eigenvalues below a scale of  $\sim \Lambda_{QCD}/L^2$
are given by chRMT. We conclude that if partially quenched chiral
perturbation theory is the correct description of the low energy limit of
QCD, we have shown that below the Thouless energy, the eigenvalue
correlations are given by chRMT. Our confidence in these arguments has
been strengthened by universality studies. 
Universal features in Dirac  spectra at nonzero chemical potential
have been identified. 

\section*{Acknowledgments}

This work was partially supported by the US DOE grant
DE-FG-88ER40388. The TPI at Minneapolis is thanked for its hospitality.
Useful discussions with B. Altshuler, P. Damgaard, T. Guhr, Y. Fyodorov, 
B. Shklovskii, A. Smilga, H. Weidenm\"uller and 
T. Wettig  are acknowledged. 
D. Toublan is thanked for a critical reading of the manuscript. 
Finally, I thank my collaborators on whose work this review is based.

\end{document}